\documentclass[twocolumn,showpacs,prl,aps,floatfix,reprint]{revtex4-1}
\usepackage{amssymb}
\usepackage{amsmath}
\usepackage{graphicx}
\usepackage{dcolumn}
\usepackage{amsfonts}
\usepackage{bm}
\usepackage[dvipsnames]{color}

\def \G{\mathcal{G}}
\def \T{\mathcal{T}}

\def \A{\mathcal{A}}

\begin{document}

\title{Pairing in high-density neutron matter including short- and long-range correlations}
\author{D. Ding$^{1}$}
\author{A. Rios$^2$}

\author{W. H. Dickhoff$^{1}$}
\author{H. Dussan$^{1}$}
\author{A. Polls$^{3}$}
\author{S. J. Witte$^{1,4}$}
\affiliation{${}^1$Department of Physics, Washington University, St.
Louis, Missouri 63130, USA}
\affiliation{${}^2$Department of Physics, Faculty of Engineering and Physical Sciences, 
University of Surrey, 
Guildford, Surrey GU2 7XH, United Kingdom}
\affiliation{${}^3$ Departament d'Estructura i Constituents de la Mat\`{e}ria and Institut de Ci\`{e}nces del Cosmos, Universitat de Barcelona, Avenida Diagonal 647, E-8028 Barcelona, Spain}
\affiliation{${}^4$Department of Physics and Astronomy,
University of California, Los Angeles, CA 90095, USA}

\date{\today}

\begin{abstract}
The influence of short-range correlations (SRC) on the spectral distribution of neutrons is incorporated in the solution of the gap equation for the ${}^3P_2-{}^3F_2$ coupled channel in pure neutron matter at high density.
This effect is studied for three different realistic interactions.
The gap in this channel is strongly suppressed by these correlations but does not vanish.
For a consistent treatment we also include for the first time the effect of long-range correlations (LRC) by incorporating polarization terms in addition to the bare interaction. This allows the neutrons to exchange density and spin fluctuations governed by the strength of Landau parameters with values that are consistent with the available literature.
While these LRC have an antiscreening tendency, they only slightly increase the gap in the ${}^3P_2-{}^3F_2$ coupled channel for all three realistic interactions as long as SRC are included.
All three interactions generate maximum gaps around 0.1 to 0.2 MeV at most with a small dependence on the hardness of the interaction. 
These results are relevant for the cooling scenarios of neutron stars, in particular the young neutron star in Cassiopeia A.
\end{abstract}
\pacs{21.65.-f, 21.65.Cd, 24.10.Cn, 26.60.-c, 74.20.Fg}
\maketitle

Pulsar glitches are considered prime evidence of superfluid properties in neutron stars on account of their long relaxation times~\cite{Migdal59,Chamel08}.
The cooling scenario of a neutron star also depends sensitively on the possible superfluidity of neutrons and superconductivity of protons in the interior of the system~\cite{Page2004}.
Recent observations of a young neutron star in Cassiopeia A~\cite{Ho09} and its rapid cooling~\cite{Shternin11} have rekindled interest in high-density neutron superfluidity~\cite{Page11,Blaschke13,Ho2015}.
Several scenarios with apparently different mechanisms appear to be able to explain the observed cooling observation~\cite{Page11,Blaschke13,Ho2015} but all require a contribution of neutron pairing with ${}^3P_2-^3F_2$ quantum numbers.
Longer observation of this system may shed further light on this issue.
Meanwhile, it appears worthwhile to consider the present status of high-density neutron pairing.

At low density there is universal agreement that neutron pairs with ${}^1S_0$ quantum numbers will be superfluid~\cite{Dean03,Gandolfi2009} on account of the exhibited attraction of the corresponding low-energy phase shift.
At higher energy, the $P$-wave phase shift of the coupled ${}^3P_2-^3F_2$ channel exhibits more attraction and therefore becomes a candidate for neutron pairing at higher density~\cite{Dean03}.
While ${}^1S_0$ gaps in standard BCS calculations are quite similar for different realistic interactions,
there is more variation in the gap obtained for the ${}^3P_2-^3F_2$ channel.
Depending on the choice of the nucleon-nucleon (NN) interaction, the results range from about 0.5 to 0.9 MeV for the maximum gap with a density range associated with Fermi momenta from 1.8 to 2.5 fm$^{-1}$~\cite{Baldo1998,Dean03,Khodel2004}.

Another example of an attractive interaction pertaining to a coupled channel occurs in symmetric nuclear matter for isoscalar pairs with ${}^3S_1$-${}^3D_1$ quantum numbers.
We build on the experience with pairing in this coupled channel to motivate our treatment of neutron ${}^3P_2-^3F_2$ pairing.
We note that early calculations around normal nuclear matter density generated a sizable gap of around 10 MeV~\cite{ars:90,vgdpr:91,Baldo92,Takatsuka93,Baldo95} with ${}^3S_1$-${}^3D_1$ quantum numbers 
when employing \textit{e.g.} Brueckner-Hartree-Fock (BHF) single-particle (sp) energies.
The empirical data of finite nuclei give no indication for
such strong proton-neutron pairing correlations with corresponding gaps as large as 10 MeV. 
It is therefore plausible that the evaluation of the ${}^3S_1$-${}^3D_1$ gaps in infinite matter according to Refs.~\cite{ars:90,vgdpr:91,Baldo92,Takatsuka93,Baldo95} suffers from the lack of inclusion of relevant physics that can suppress such a large gap.

Nucleons in nuclei are well documented to behave differently than expected on the basis of a pure independent particle model (IPM).
Additional correlations beyond the Pauli principle of the IPM are clearly established in the analysis of the $(e,e'p)$ reaction of closed-shell nuclei~\cite{Lapikas93} leading to removal probabilities for valence protons that are reduced by about 0.35 from the IPM value of 1.
A substantial contribution to this reduction, about 0.10 to 0.15, is a consequence of the depletion of the Fermi sea due to short-range correlations (SRC)~\cite{Dickhoff04}.
The complementary admixture of a modest amount of high-momentum components in the nuclear ground state  has been experimentally observed~\cite{Rohe2004}. 

Such correlations can be accurately and completely treated in infinite nuclear systems where the consequences of summing ladder diagrams can be included into the self-energy and off-shell propagation can be incorporated self-consistently~\cite{Dickhoff04}.
While such a procedure still encounters some numerical issues at zero temperature no such problems are encountered at sufficiently high temperature~\cite{libphd,frickphd,riosphd,Dickhoff2008}.
At finite temperature it is thus possible to fully account for the influence of SRC on the propagation properties of nucleons in symmetric nuclear and neutron matter.

At $T = 0$ ladder-diagram summations can lead to pairing instabilities when the possibility of anomalous (pair)  propagation is not included.
Since the influence of pairing on the normal nucleon propagator is confined to energies around the Fermi energy that are of the order of the size of the gap, one may reasonably assume that the normal self-energy will hardly be affected by pairing correlations~\cite{Migdal1967}.
Making this assumption and extrapolating the temperature dependence of the self-consistently calculated normal self-energy to lower temperatures including $T=0$, it was shown in Ref.~\cite{Muther2005} that due to the influence of SRC  the 
${}^3S_1$-${}^3D_1$ gap at normal density vanishes, reconciling empirical information with many-body calculations of pairing in nuclear matter.
We therefore follow such a procedure for the calculation of the influence of SRC on the pairing properties in neutron matter at high density for the ${}^3P_2-^3F_2$ coupled channel.
 
 Another important ingredient for the calculation of pairing correlations at high density in neutron matter is the modification of the pairing interaction when nucleons exchange the low-energy, possibly collective, excitations of the medium.
At small momentum transfer such excitations are determined by the Landau parameters that govern the density and spin excitations of the medium~\cite{Dickhoff2008}.
The first correction to be added to the bare NN interaction is given by the exchange of one particle-hole bubble between the neutrons.
This mechanism can be extended to exchanging an infinite-order random-phase-approximation (RPA) bubble series.
Such contributions arise since the gap equation itself generates the contribution of ladder diagrams~\cite{Migdal1967,Dickhoff88,Dickhoff2008}.
The accurate calculation of Landau parameters depends on many ingredients some of which were discussed in Ref.~\cite{Dickhoff1987}.
We will also include the contribution of these so-called polarization terms that reflect LRC and follow the procedure outlined in Ref.~\cite{Cao2006} for neutron ${}^1S_0$ superfluidity and extend it to the case of the ${}^3P_2-^3F_2$ coupled partial waves thereby generating for the first time a complete treatment of both SRC and LRC for this pairing channel.
 
We start with a brief and schematic reminder of the self-consistent summation of ladder diagrams that was used for the construction of the ingredients of the pairing calculation~\cite{Rios2009a,Rios2014}.
The proper treatment of SRC and tensor correlations in the nuclear medium is accomplished by constructing the
in-medium Lippman-Schwinger equation, schematically represented by:
\begin{equation}
	\T = V + V \G_{II}^0 \T \, .
\label{eq:LSeq}
\end{equation}
The NN interaction is denoted by $V$ and the noninteracting but dressed two-particle propagator is given by
the convolution of two dressed sp propagators
\begin{eqnarray}
\G_{II}^0(k,k'; \Omega) & = & 
\int \frac{\textrm{d} \omega}{2 \pi} \frac{\textrm{d} \omega'}{2 \pi}  \A(k,\omega) \A(k',\omega') \nonumber \\
&& \times \frac{ 1-f(\omega)-f(\omega')}{\Omega - \omega - \omega' +i\eta} \, ,
\label{eq:g2}
\end{eqnarray}
where $f(\omega) = ( 1 + e^{  ( \omega - \mu )/T} )^{-1}$ is the Fermi-Dirac distribution and $\A$ the sp spectral function.
The temperature, $T$, is applied as an external condition. 
The chemical potential, $\mu$, is obtained from the normalization
to the density of the momentum distribution.
Self-consistency is imposed all the way through in our calculations. The 
sp spectral functions
that enter Eq.~(\ref{eq:g2}), for instance, are obtained from the $\T-$matrix itself. 
In the ladder approximation, 
this effective interaction determines the imaginary part of the self-energy~\cite{Rios2009a,Rios2014}.

The dispersive contribution to the real part of the self-energy can be obtained from
a dispersion relation~\cite{Dickhoff2008}. 
In addition the energy-independent correlated Hartree-Fock contribution involving the actual momentum distribution
must be included as well. 
From Dyson's equation we then obtain the sp spectral function:
\begin{equation}
\A(k,\omega) = \frac{-2 \textrm{Im} \Sigma(k,\omega) }
{ [ \omega - \frac{k^2}{2m} - \textrm{Re} \Sigma(k,\omega) ]^2+ [ \textrm{Im} \Sigma(k,\omega) ]^2} \, ,
\label{eq:asf}
\end{equation}
which closes the self-consistency loop that treats all particles in the same manner, 
providing appropriate feedback to the different ingredients of the calculations.

The proper treatment of pairing requires the introduction of anomalous propagators~\cite{Migdal1967,Dickhoff2008} with corresponding anomalous self-energy terms in so-called Gorkov equations~\cite{Gorkov58,Migdal1967,Barbieri2011} that couple normal and abnormal (pair) propagators.
The resulting expression for the anomalous self-energy $\Delta$ can be written in the form of the usual gap equation in a partial wave expansion~\cite{Bozek99,Muther2005}
\begin{equation}
\Delta^{JST}_\ell(p) = - \sum_{\ell'} \int_0^{\infty} 
\frac{{\mathrm{d}} k\,k^{2}}{(2\pi)^3} \left<p|V^{JST}_{\ell \ell'}|k\right>_A \frac{1}{2E_k}
\Delta^{JST}_{\ell'}(k) \,.
\label{eq:gapbcs}
\end{equation}
For a coupled channel we employ the spherical gap
\begin{equation}
\Delta^{JST}(k) = \left[\left(\Delta^{JST}_{\ell=J-1}(k)\right)^2+\left(\Delta^{JST}_{\ell=J+1}(k)\right)^2\right]^{1/2}
\label{eq:gapav}
\end{equation}
in $E_k = \sqrt{\chi_k^2 + [\Delta^{JST}(k)]^2}$. With $\chi_k=\mu -\varepsilon_k$, while $\varepsilon_k$ represents the kinetic energy or medium-modified spectrum without pairing, one obtains the conventional BCS gap.
The inclusion of SRC requires the calculation of
\begin{eqnarray}
-\frac{1}{2E_k}\quad & \to & \frac{-1}{2 \sqrt{\widetilde{\chi}_k^2 +  [\Delta^{JST}(k)]^2}} = 
\int \frac{{\mathrm{d}}\omega}{2\pi} \frac{{\mathrm{d}}\omega^{\prime}}{2\pi} 
 \nonumber \\
&& \times \A(k,\omega) \A_s(k,\omega^{\prime})
\frac{1-f(\omega)-f(\omega^{\prime})}
{-\omega-\omega^{\prime}}, \,\label{def:zk0}
\end{eqnarray}
%
where $\A_s$ is the spectral function that includes a nontrivial pairing solution to the gap equation.
It is therefore appropriate to consider Eq.~(\ref{eq:gapbcs}) as a generalization of the usual
gap equation when $E_k$ is obtained from Eq.~(\ref{def:zk0}), as it accounts for the spreading of sp strength due to SRC and tensor correlations~\cite{Dewulf2003,Rios2009a}.
For practical calculations we assume that different partial-wave projections do not couple through the angular dependence of the interaction and the propagators. 
We therefore employ the spherical version of the quasiparticle energy as in Eq.~(\ref{eq:gapav}).

The essential calculation that provides the input for the solution of the gap equation of Eq.~(\ref{eq:gapbcs}) is provided by a proper temperature extrapolation~\cite{Suppl15} of the convolution of spectral functions generated by the self-consistent normal self-energy in which $\A_s$ in Eq.~(\ref{def:zk0}) is replaced by $\A$ whereby we obtain a generalized version $\widetilde{\chi}_k$ instead of $\chi_k$~\cite{Muther2005} 
at a given temperature.
For the present work we have explored an expansion of both the real and imaginary part of the normal self-energy in even powers of the temperature.
This expansion is constrained by relevant macroscopic thermodynamical quantities.
The zero-temperature limit can then be used to construct the relevant spectral functions modified by SRC.
This procedure was tested by comparing with the results of Ref.~\cite{Muther2005} for nuclear and neutron matter in which a slightly different extrapolation method was employed. 

We also report results for the inclusion of polarization terms in the pairing interaction.
We have followed the procedure proposed in Ref.~\cite{Cao2006} and generalized it to the case of the ${}^3P_2-^3F_2$ coupled channel.
The physical exchange of low-lying possibly collective density and spin modes can be included by adding the corresponding interaction terms to the bare NN interaction.
These excitations are dominated by the strength of the corresponding Landau parameters which govern the low-lying spectrum of the fluid. 
The possibility that the presence of the pion-exchange tensor interaction strongly influences the spin mode~\cite{Dickhoff1982} cannot be excluded but is beyond the scope of the present work~\cite{Pankratov2015}. 
The corresponding interaction that treats LRC for the $^1S_0$ channel is given by
\begin{equation}\label{V2-1S0}
V_{LRC}^{S=0}(q)=\frac{1}{2}G_{ph}^0G_{ph}^0\Lambda^0(q)-\frac{3}{2}G_{ph}^1G_{ph}^1\Lambda^1(q),
\end{equation}
where $G_{ph}^{S_{ph}}$ are particle-hole transformed G-matrix elements averaged around the Fermi energy~\cite{Cao2006}.
The iterated bubble series is represented by
\begin{equation}
\Lambda^S(q)=\frac{\Lambda_0(q)}{1-\Lambda_0(q)*\mathcal{L}^S}. \label{eq:bubbles}
\end{equation}
The $\mathcal{L}^S$ correspond to Landau parameters. The density mode with total spin 0 in the particle-hole channel is determined by $\mathcal{L}^{0}$ which is attractive at low density and the spin mode with total spin 1 by $\mathcal{L}^{1}$ which is repulsive but has similar magnitude, while both exhibit a modest density dependence~\cite{Cao2006}.
The static Lindhard function $\Lambda_0(q)$ is iterated to all orders in this procedure to generate the 
$\Lambda^S$ which are negative. The first term in Eq.~(\ref{V2-1S0}) is attractive and the second term is repulsive  but dominates on account of the spin factor ultimately leading to additional suppression of the gap in this channel.

For the $^3P_2-^3F_2$ channel which involves spin-1 pairs, the sampling over density and spin modes yields
\begin{equation}\label{V2-3pf}
V_{LRC}^{S=1}(q)=\frac{1}{2}G_{ph}^0G_{ph}^0\Lambda^0(q)+\frac{1}{2}G_{ph}^1G_{ph}^1\Lambda^1(q), 
\end{equation}
with both terms yielding attraction.
Contrary to the ${}^1S_0$ channel this contribution leads to antiscreening of the gap as it represents an attractive interaction.
The Landau parameters  $\mathcal{L}^S$ from Ref.~\cite{Cao2006} are extrapolated to higher densities in a smooth way to approximately fulfill the forward scattering sum rule for these parameters~\cite{Friman1979,Dickhoff1987} leading to their sign reversal at a density corresponding to about 1.8 fm$^{-1}$.

\begin{figure}[b]
\includegraphics[height=0.3\textwidth]{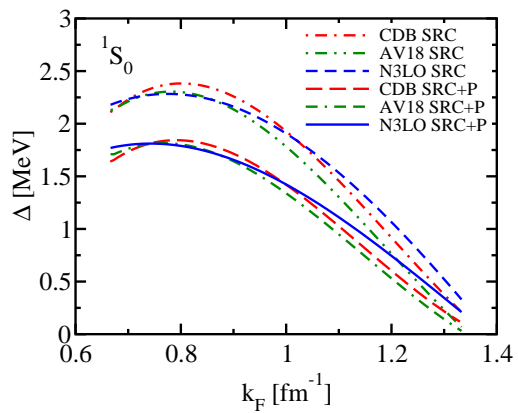}
 \caption{Results for the ${}^1S_0$ gap in neutron matter for three different realistic interaction with the inclusion of SRC and separately with additional polarization terms. }
\label{fig:1S0}
\end{figure}

\begin{figure*}[bht]
\includegraphics[height=.22\textheight]{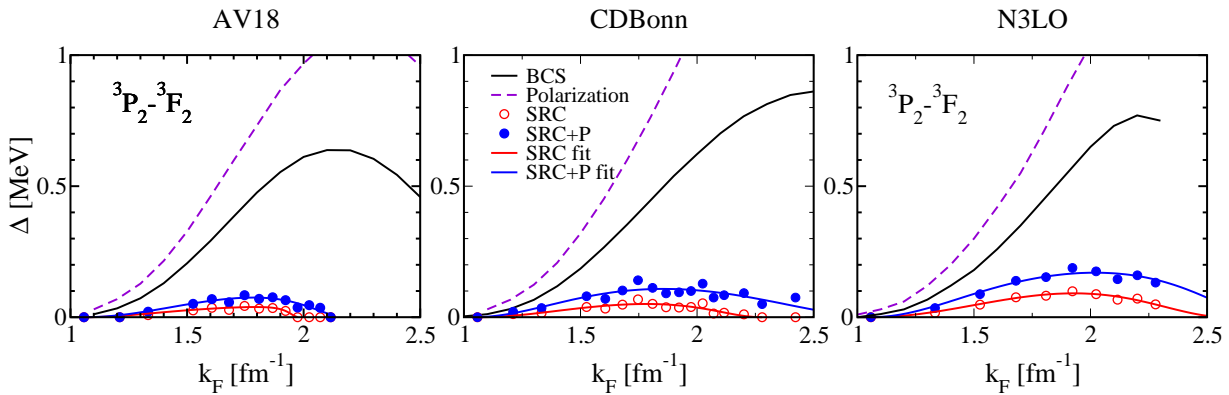}
\caption{Gap for ${}^3P_2-^3F_2$ neutron pairing as a function of the Fermi momentum in pure neutron matter using the AV18 (left panel), CDBonn (middle panel), and N3LO interaction (right panel). For symbols see the text.}
\label{fig:gapAV18}
\end{figure*}

Results for the ${}^1S_0$ channel including SRC are consistent with the results of Ref.~\cite{Muther2005} for the charge-dependent Bonn (CDBonn) interaction~\cite{Machleidt1995} and Argonne V18 (AV18) interaction of Ref.~\cite{Wiringa1995}.
The inclusion of polarization terms on top of the bare interaction also reproduce the results of Ref.~\cite{Cao2006}. 
The main effect of the inclusion of SRC in this channel is to reduce the value of the gap by about 0.75 MeV compared to the BCS results that employ a free sp spectrum and generate a gap closure at a corresponding lower density.
We note that the inclusion of self-energy effects due to SRC removes sp strength from the vicinity of the Fermi energy and distributes it far and wide without compensating the loss of strength near the Fermi energy.
This feature has the inescapable consequence that a pairing solution is less likely and therefore the gap is suitably reduced or vanishes.
Other calculations that take such effects into account yield identical conclusions~\cite{Bozek00,Baldo00,Baldo02,Bozek03}.

Results for the Idaho next-to-next-to-next-to-leading-order
(N3LO) chiral-perturbation-theory potential of Ref.~\cite{Entem2003} have also been generated.
Our results for the three realistic interactions including both SRC and LRC are very similar and characterized by a maximum gap of about 1.8 MeV at a density corresponding to a Fermi momentum of a little more than 0.8 fm${}^{-1}$.
These results appear very robust and confirm the general understanding of the gap in this channel in terms of its size and density dependence~\cite{Gandolfi2009}.
Results are shown in Fig.~\ref{fig:1S0} for the CDBonn interaction by the dash-dot (SRC only) and long-dashed curve (SRC + polarization), N3LO curves by the short-dashed (SRC) and  solid curve (SRC + polarization), and AV18 curves by the dash-dot-dot (SRC) and dash-dah-dot curve (SRC + polarization).

In the case of the ${}^3P_2-^3F_2$ channel, the spin recoupling from particle-hole to particle-particle leads to an interaction that enhances the pairing gap and can therefore be characterized as antiscreening according to Eq.~({\ref{V2-3pf}).
Preliminary results reported in~\cite{Ding2014} generated no pairing solution in this channel.
A more detailed treatment of the extrapolation to $T=0$ has recently been implemented~\cite{Suppl15} yielding a small but nonvanishing gap for all three interactions.
We show in Fig.~\ref{fig:gapAV18} the results for these interactions including the standard BCS result (solid line), additional inclusion of polarization (dashed), inclusion of SRC (open circles), and inclusion of both SRC and LRC (full circles). 
A fit to the calculated data points is represented by solid curves using the parametrization of Ref.~\cite{Ho2015} (see the table in Ref.~\cite{Suppl15}).
The calculations for the N3LO interaction have been restricted to densities somewhat below its 500 MeV cut-off~\cite{Maurizio2014}.

The BCS results for these interactions are quite similar up to a density corresponding to $k_F \approx 2.0\ \textrm{fm}^{-1}$ but slightly differ in their maximum gap and the density for gap closure. 
CDBonn generates the largest values for these quantities.
The effect of adding only polarization to the BCS treatment then yields dramatically different results illustrated by the dashed curves in Fig.~\ref{fig:gapAV18} with maxima (not shown) corresponding to 1.46 (AV18) , 2.68 (CDBonn), and 1.33 MeV (N3LO), respectively.
When only the SRC are incorporated the results for the three interactions indicated by open circles become much closer with maximum gaps ranging from 0.04 MeV (AV18), 0.07 (CDBonn) to a about 0.1 MeV (N3LO).
The size of the employed symbols in the figures reflect a reasonable estimate of the error associated with the $T=0$ extrapolation~\cite{Suppl15}.
The additional inclusion of LRC to the SRC effect enhances the gap slightly for all three interactions with the softest interaction (N3LO) generating the largest gap of about 0.2 MeV, CDBonn 0.1 MeV, and the harder AV18 a maximum of about 0.08 MeV.


We conclude that the inclusion of SRC has a dramatic but not unexpected effect on the solution of the gap equation in high-density neutron matter for the ${}^3P_2-^3F_2$ coupled channel.
This conclusion is valid for the hard AV18, the softer CDBonn, as well as the very soft N3LO interaction.
Our results indicate that a small gap remains at high density irrespective of the realistic interaction. The treatment of LRC suggests that SRC generate the dominant effect but they do enhance the gap to maximum values ranging from 0.1 to 0.2 MeV at densities corresponding to  $k_F \approx 1.7\ \textrm{fm}^{-1}$ and $k_F \approx 1.9\ \textrm{fm}^{-1}$, respectively.
More sophisticated treatments of LRC~\cite{Sedrakian03} appear unlikely to change these results by much as the effect of SRC remains dominant.
We note that the smallness of the gaps may require an accuracy that may be difficult to reach by Monte Carlo calculations in the near future.
The inclusion of three-body forces associated with the N3LO interaction appears possible~\cite{Carbone2013a,carbonephd} but may not change these conclusions according to Ref.~\cite{Dong2013}.

The present results yield for the first time a consistent treatment of both SRC and LRC for the ${}^3P_2-^3F_2$ coupled channel and suggest that a small gap in a narrow density regime can be expected at densities where realistic interactions are still constrained by NN scattering data.
The impact of our conclusions on cooling
calculations is likely to be qualitatively significant. 
A complete understanding of cooling phenomena requires a consistent description of both the equation of state as well as pairing properties.
The present effort represents a first step in providing these ingredients.


This material is based upon work supported by the U.S. National Science Foundation under grant PHY-1304242,
Consolider
Ingenio 2010 Programme CPAN CSD2007-00042, Grant
No. FIS2011-24154 from MICINN (Spain), and Grant No.
2014SGR-401 from Generalitat de Catalunya (Spain); by
STFC, through GrantsNo. ST/I005528/1 and No. ST/J000051; and by NewCompstar, COST Action MP1304.

\bibliography{biblio3PF2}

\begin{thebibliography}{52}%
\makeatletter
\providecommand \@ifxundefined [1]{%
 \@ifx{#1\undefined}
}%
\providecommand \@ifnum [1]{%
 \ifnum #1\expandafter \@firstoftwo
 \else \expandafter \@secondoftwo
 \fi
}%
\providecommand \@ifx [1]{%
 \ifx #1\expandafter \@firstoftwo
 \else \expandafter \@secondoftwo
 \fi
}%
\providecommand \natexlab [1]{#1}%
\providecommand \enquote  [1]{``#1''}%
\providecommand \bibnamefont  [1]{#1}%
\providecommand \bibfnamefont [1]{#1}%
\providecommand \citenamefont [1]{#1}%
\providecommand \href@noop [0]{\@secondoftwo}%
\providecommand \href [0]{\begingroup \@sanitize@url \@href}%
\providecommand \@href[1]{\@@startlink{#1}\@@href}%
\providecommand \@@href[1]{\endgroup#1\@@endlink}%
\providecommand \@sanitize@url [0]{\catcode `\\12\catcode `\$12\catcode
  `\&12\catcode `\#12\catcode `\^12\catcode `\_12\catcode `\%12\relax}%
\providecommand \@@startlink[1]{}%
\providecommand \@@endlink[0]{}%
\providecommand \url  [0]{\begingroup\@sanitize@url \@url }%
\providecommand \@url [1]{\endgroup\@href {#1}{\urlprefix }}%
\providecommand \urlprefix  [0]{URL }%
\providecommand \Eprint [0]{\href }%
\providecommand \doibase [0]{http://dx.doi.org/}%
\providecommand \selectlanguage [0]{\@gobble}%
\providecommand \bibinfo  [0]{\@secondoftwo}%
\providecommand \bibfield  [0]{\@secondoftwo}%
\providecommand \translation [1]{[#1]}%
\providecommand \BibitemOpen [0]{}%
\providecommand \bibitemStop [0]{}%
\providecommand \bibitemNoStop [0]{.\EOS\space}%
\providecommand \EOS [0]{\spacefactor3000\relax}%
\providecommand \BibitemShut  [1]{\csname bibitem#1\endcsname}%
\let\auto@bib@innerbib\@empty
\bibitem [{\citenamefont {Migdal}(1959)}]{Migdal59}%
  \BibitemOpen
  \bibfield  {author} {\bibinfo {author} {\bibfnamefont {A.~B.}\ \bibnamefont
  {Migdal}},\ }\href@noop {} {\bibfield  {journal} {\bibinfo  {journal} {Nucl.
  Phys.}\ }\textbf {\bibinfo {volume} {13}},\ \bibinfo {pages} {655} (\bibinfo
  {year} {1959})}\BibitemShut {NoStop}%
\bibitem [{\citenamefont {Chamel}\ and\ \citenamefont
  {Haensel}(2008)}]{Chamel08}%
  \BibitemOpen
  \bibfield  {author} {\bibinfo {author} {\bibfnamefont {N.}~\bibnamefont
  {Chamel}}\ and\ \bibinfo {author} {\bibfnamefont {P.}~\bibnamefont
  {Haensel}},\ }\href@noop {} {\bibfield  {journal} {\bibinfo  {journal}
  {Living Rev. Relativity}\ }\textbf {\bibinfo {volume} {11}},\ \bibinfo
  {pages} {10} (\bibinfo {year} {2008})}\BibitemShut {NoStop}%
\bibitem [{\citenamefont {Page}\ \emph {et~al.}(2004)\citenamefont {Page},
  \citenamefont {Lattimer}, \citenamefont {Prakash},\ and\ \citenamefont
  {Steiner}}]{Page2004}%
  \BibitemOpen
  \bibfield  {author} {\bibinfo {author} {\bibfnamefont {D.}~\bibnamefont
  {Page}}, \bibinfo {author} {\bibfnamefont {J.~M.}\ \bibnamefont {Lattimer}},
  \bibinfo {author} {\bibfnamefont {M.}~\bibnamefont {Prakash}}, \ and\
  \bibinfo {author} {\bibfnamefont {A.~W.}\ \bibnamefont {Steiner}},\
  }\href@noop {} {\bibfield  {journal} {\bibinfo  {journal} {Astrophys. J.
  Suppl. Ser.}\ }\textbf {\bibinfo {volume} {155}},\ \bibinfo {pages} {623}
  (\bibinfo {year} {2004})}\BibitemShut {NoStop}%
\bibitem [{\citenamefont {Ho}\ and\ \citenamefont {Heinke}(2009)}]{Ho09}%
  \BibitemOpen
  \bibfield  {author} {\bibinfo {author} {\bibfnamefont {W.~C.~G.}\
  \bibnamefont {Ho}}\ and\ \bibinfo {author} {\bibfnamefont {C.~O.}\
  \bibnamefont {Heinke}},\ }\href@noop {} {\bibfield  {journal} {\bibinfo
  {journal} {Nature}\ }\textbf {\bibinfo {volume} {492}},\ \bibinfo {pages}
  {71} (\bibinfo {year} {2009})}\BibitemShut {NoStop}%
\bibitem [{\citenamefont {Shternin}\ \emph {et~al.}(2011)\citenamefont
  {Shternin}, \citenamefont {Yakovlev}, \citenamefont {Heinke}, \citenamefont
  {Ho},\ and\ \citenamefont {Patnaude}}]{Shternin11}%
  \BibitemOpen
  \bibfield  {author} {\bibinfo {author} {\bibfnamefont {P.~S.}\ \bibnamefont
  {Shternin}}, \bibinfo {author} {\bibfnamefont {D.~G.}\ \bibnamefont
  {Yakovlev}}, \bibinfo {author} {\bibfnamefont {C.~O.}\ \bibnamefont
  {Heinke}}, \bibinfo {author} {\bibfnamefont {W.~C.~G.}\ \bibnamefont {Ho}}, \
  and\ \bibinfo {author} {\bibfnamefont {D.~J.}\ \bibnamefont {Patnaude}},\
  }\href@noop {} {\bibfield  {journal} {\bibinfo  {journal} {Mom. Not. R.
  Astron. Soc.}\ }\textbf {\bibinfo {volume} {412}},\ \bibinfo {pages} {L108}
  (\bibinfo {year} {2011})}\BibitemShut {NoStop}%
\bibitem [{\citenamefont {Page}\ \emph {et~al.}(2011)\citenamefont {Page},
  \citenamefont {Prakash}, \citenamefont {Lattimer},\ and\ \citenamefont
  {Steiner}}]{Page11}%
  \BibitemOpen
  \bibfield  {author} {\bibinfo {author} {\bibfnamefont {D.}~\bibnamefont
  {Page}}, \bibinfo {author} {\bibfnamefont {M.}~\bibnamefont {Prakash}},
  \bibinfo {author} {\bibfnamefont {J.~M.}\ \bibnamefont {Lattimer}}, \ and\
  \bibinfo {author} {\bibfnamefont {A.~W.}\ \bibnamefont {Steiner}},\
  }\href@noop {} {\bibfield  {journal} {\bibinfo  {journal} {Phys. Rev. Lett.}\
  }\textbf {\bibinfo {volume} {106}},\ \bibinfo {pages} {081101} (\bibinfo
  {year} {2011})}\BibitemShut {NoStop}%
\bibitem [{\citenamefont {Blaschke}\ \emph {et~al.}(2013)\citenamefont
  {Blaschke}, \citenamefont {Grigorian},\ and\ \citenamefont
  {Voskresensky}}]{Blaschke13}%
  \BibitemOpen
  \bibfield  {author} {\bibinfo {author} {\bibfnamefont {D.}~\bibnamefont
  {Blaschke}}, \bibinfo {author} {\bibfnamefont {H.}~\bibnamefont {Grigorian}},
  \ and\ \bibinfo {author} {\bibfnamefont {D.~N.}\ \bibnamefont
  {Voskresensky}},\ }\href@noop {} {\bibfield  {journal} {\bibinfo  {journal}
  {Phys. Rev. C}\ }\textbf {\bibinfo {volume} {88}},\ \bibinfo {pages} {065805}
  (\bibinfo {year} {2013})}\BibitemShut {NoStop}%
\bibitem [{\citenamefont {Ho}\ \emph {et~al.}(2015)\citenamefont {Ho},
  \citenamefont {Elshamouty}, \citenamefont {Heinke},\ and\ \citenamefont
  {Potekhin}}]{Ho2015}%
  \BibitemOpen
  \bibfield  {author} {\bibinfo {author} {\bibfnamefont {W.~C.~G.}\
  \bibnamefont {Ho}}, \bibinfo {author} {\bibfnamefont {K.~G.}\ \bibnamefont
  {Elshamouty}}, \bibinfo {author} {\bibfnamefont {C.~O.}\ \bibnamefont
  {Heinke}}, \ and\ \bibinfo {author} {\bibfnamefont {A.~Y.}\ \bibnamefont
  {Potekhin}},\ }\href@noop {} {\bibfield  {journal} {\bibinfo  {journal}
  {Phys. Rev. C}\ }\textbf {\bibinfo {volume} {91}},\ \bibinfo {pages} {015806}
  (\bibinfo {year} {2015})}\BibitemShut {NoStop}%
\bibitem [{\citenamefont {Dean}\ and\ \citenamefont
  {Hjorth-Jensen}(2003)}]{Dean03}%
  \BibitemOpen
  \bibfield  {author} {\bibinfo {author} {\bibfnamefont {D.~J.}\ \bibnamefont
  {Dean}}\ and\ \bibinfo {author} {\bibfnamefont {M.}~\bibnamefont
  {Hjorth-Jensen}},\ }\href@noop {} {\bibfield  {journal} {\bibinfo  {journal}
  {Rev. Mod. Phys.}\ }\textbf {\bibinfo {volume} {75}},\ \bibinfo {pages} {607}
  (\bibinfo {year} {2003})}\BibitemShut {NoStop}%
\bibitem [{\citenamefont {Gandolfi}\ \emph {et~al.}(2009)\citenamefont
  {Gandolfi}, \citenamefont {Illarionov}, \citenamefont {Pederiva},
  \citenamefont {Schmidt},\ and\ \citenamefont {Fantoni}}]{Gandolfi2009}%
  \BibitemOpen
  \bibfield  {author} {\bibinfo {author} {\bibfnamefont {S.}~\bibnamefont
  {Gandolfi}}, \bibinfo {author} {\bibfnamefont {A.~Y.}\ \bibnamefont
  {Illarionov}}, \bibinfo {author} {\bibfnamefont {F.}~\bibnamefont
  {Pederiva}}, \bibinfo {author} {\bibfnamefont {K.~E.}\ \bibnamefont
  {Schmidt}}, \ and\ \bibinfo {author} {\bibfnamefont {S.}~\bibnamefont
  {Fantoni}},\ }\href@noop {} {\bibfield  {journal} {\bibinfo  {journal} {Phys.
  Rev. C}\ }\textbf {\bibinfo {volume} {80}},\ \bibinfo {pages} {045802}
  (\bibinfo {year} {2009})}\BibitemShut {NoStop}%
\bibitem [{\citenamefont {Baldo}\ \emph {et~al.}(1998)\citenamefont {Baldo},
  \citenamefont {Elgaroy}, \citenamefont {Engvik}, \citenamefont
  {{Hjorth-Jensen}},\ and\ \citenamefont {Schulze}}]{Baldo1998}%
  \BibitemOpen
  \bibfield  {author} {\bibinfo {author} {\bibfnamefont {M.}~\bibnamefont
  {Baldo}}, \bibinfo {author} {\bibfnamefont {O.}~\bibnamefont {Elgaroy}},
  \bibinfo {author} {\bibfnamefont {L.}~\bibnamefont {Engvik}}, \bibinfo
  {author} {\bibfnamefont {M.}~\bibnamefont {{Hjorth-Jensen}}}, \ and\ \bibinfo
  {author} {\bibfnamefont {H.~J.}\ \bibnamefont {Schulze}},\ }\href@noop {}
  {\bibfield  {journal} {\bibinfo  {journal} {Phys. Rev. C}\ }\textbf {\bibinfo
  {volume} {58}},\ \bibinfo {pages} {1921} (\bibinfo {year}
  {1998})}\BibitemShut {NoStop}%
\bibitem [{\citenamefont {Khodel}\ \emph {et~al.}(2004)\citenamefont {Khodel},
  \citenamefont {Clark}, \citenamefont {Takano},\ and\ \citenamefont
  {Zverev}}]{Khodel2004}%
  \BibitemOpen
  \bibfield  {author} {\bibinfo {author} {\bibfnamefont {V.~A.}\ \bibnamefont
  {Khodel}}, \bibinfo {author} {\bibfnamefont {J.~W.}\ \bibnamefont {Clark}},
  \bibinfo {author} {\bibfnamefont {M.}~\bibnamefont {Takano}}, \ and\ \bibinfo
  {author} {\bibfnamefont {M.~V.}\ \bibnamefont {Zverev}},\ }\href@noop {}
  {\bibfield  {journal} {\bibinfo  {journal} {Phys. Rev. Lett.}\ }\textbf
  {\bibinfo {volume} {93}},\ \bibinfo {pages} {151101} (\bibinfo {year}
  {2004})}\BibitemShut {NoStop}%
\bibitem [{\citenamefont {Alm}\ \emph {et~al.}(1990)\citenamefont {Alm},
  \citenamefont {R{\"{o}}pke},\ and\ \citenamefont {Schmidt}}]{ars:90}%
  \BibitemOpen
  \bibfield  {author} {\bibinfo {author} {\bibfnamefont {T.}~\bibnamefont
  {Alm}}, \bibinfo {author} {\bibfnamefont {G.}~\bibnamefont {R{\"{o}}pke}}, \
  and\ \bibinfo {author} {\bibfnamefont {M.}~\bibnamefont {Schmidt}},\
  }\href@noop {} {\bibfield  {journal} {\bibinfo  {journal} {Z. Phys. A}\
  }\textbf {\bibinfo {volume} {337}},\ \bibinfo {pages} {355} (\bibinfo {year}
  {1990})}\BibitemShut {NoStop}%
\bibitem [{\citenamefont {Vonderfecht}\ \emph {et~al.}(1991)\citenamefont
  {Vonderfecht}, \citenamefont {Gearhart}, \citenamefont {Dickhoff},
  \citenamefont {Polls},\ and\ \citenamefont {Ramos}}]{vgdpr:91}%
  \BibitemOpen
  \bibfield  {author} {\bibinfo {author} {\bibfnamefont {B.~E.}\ \bibnamefont
  {Vonderfecht}}, \bibinfo {author} {\bibfnamefont {C.~C.}\ \bibnamefont
  {Gearhart}}, \bibinfo {author} {\bibfnamefont {W.~H.}\ \bibnamefont
  {Dickhoff}}, \bibinfo {author} {\bibfnamefont {A.}~\bibnamefont {Polls}}, \
  and\ \bibinfo {author} {\bibfnamefont {A.}~\bibnamefont {Ramos}},\
  }\href@noop {} {\bibfield  {journal} {\bibinfo  {journal} {Phys. Lett. B}\
  }\textbf {\bibinfo {volume} {253}},\ \bibinfo {pages} {1} (\bibinfo {year}
  {1991})}\BibitemShut {NoStop}%
\bibitem [{\citenamefont {Baldo}\ \emph {et~al.}(1992)\citenamefont {Baldo},
  \citenamefont {Bombaci},\ and\ \citenamefont {Lombardo}}]{Baldo92}%
  \BibitemOpen
  \bibfield  {author} {\bibinfo {author} {\bibfnamefont {M.}~\bibnamefont
  {Baldo}}, \bibinfo {author} {\bibfnamefont {I.}~\bibnamefont {Bombaci}}, \
  and\ \bibinfo {author} {\bibfnamefont {U.}~\bibnamefont {Lombardo}},\
  }\href@noop {} {\bibfield  {journal} {\bibinfo  {journal} {Phys. Lett. B}\
  }\textbf {\bibinfo {volume} {283}},\ \bibinfo {pages} {8} (\bibinfo {year}
  {1992})}\BibitemShut {NoStop}%
\bibitem [{\citenamefont {Takatsuka}\ and\ \citenamefont
  {Tamagaki}(1993)}]{Takatsuka93}%
  \BibitemOpen
  \bibfield  {author} {\bibinfo {author} {\bibfnamefont {T.}~\bibnamefont
  {Takatsuka}}\ and\ \bibinfo {author} {\bibfnamefont {R.}~\bibnamefont
  {Tamagaki}},\ }\href@noop {} {\bibfield  {journal} {\bibinfo  {journal}
  {Suppl. Prog. Theor. Phys.}\ }\textbf {\bibinfo {volume} {112}},\ \bibinfo
  {pages} {27} (\bibinfo {year} {1993})}\BibitemShut {NoStop}%
\bibitem [{\citenamefont {Baldo}\ \emph {et~al.}(1995)\citenamefont {Baldo},
  \citenamefont {Lombardo},\ and\ \citenamefont {Schuck}}]{Baldo95}%
  \BibitemOpen
  \bibfield  {author} {\bibinfo {author} {\bibfnamefont {M.}~\bibnamefont
  {Baldo}}, \bibinfo {author} {\bibfnamefont {U.}~\bibnamefont {Lombardo}}, \
  and\ \bibinfo {author} {\bibfnamefont {P.}~\bibnamefont {Schuck}},\
  }\href@noop {} {\bibfield  {journal} {\bibinfo  {journal} {Phys. Rev. C}\
  }\textbf {\bibinfo {volume} {52}},\ \bibinfo {pages} {975} (\bibinfo {year}
  {1995})}\BibitemShut {NoStop}%
\bibitem [{\citenamefont {Lapik{\'a}s}(1993)}]{Lapikas93}%
  \BibitemOpen
  \bibfield  {author} {\bibinfo {author} {\bibfnamefont {L.}~\bibnamefont
  {Lapik{\'a}s}},\ }\href {\doibase 10.1016/0375-9474(93)90630-G} {\bibfield
  {journal} {\bibinfo  {journal} {Nucl. Phys.}\ }\textbf {\bibinfo {volume}
  {A553}},\ \bibinfo {pages} {297c} (\bibinfo {year} {1993})}\BibitemShut
  {NoStop}%
\bibitem [{\citenamefont {Dickhoff}\ and\ \citenamefont
  {Barbieri}(2004)}]{Dickhoff04}%
  \BibitemOpen
  \bibfield  {author} {\bibinfo {author} {\bibfnamefont {W.~H.}\ \bibnamefont
  {Dickhoff}}\ and\ \bibinfo {author} {\bibfnamefont {C.}~\bibnamefont
  {Barbieri}},\ }\href {\doibase 10.1016/j.ppnp.2004.02.038} {\bibfield
  {journal} {\bibinfo  {journal} {Prog. Part. Nucl. Phys.}\ }\textbf {\bibinfo
  {volume} {52}},\ \bibinfo {pages} {377} (\bibinfo {year} {2004})}\BibitemShut
  {NoStop}%
\bibitem [{\citenamefont {Rohe}\ \emph {et~al.}(2004)\citenamefont {Rohe} \emph
  {et~al.}}]{Rohe2004}%
  \BibitemOpen
  \bibfield  {author} {\bibinfo {author} {\bibfnamefont {D.}~\bibnamefont
  {Rohe}} \emph {et~al.},\ }\href {\doibase 10.1103/PhysRevLett.93.182501}
  {\bibfield  {journal} {\bibinfo  {journal} {Phys. Rev. Lett.}\ }\textbf
  {\bibinfo {volume} {93}},\ \bibinfo {pages} {182501} (\bibinfo {year}
  {2004})}\BibitemShut {NoStop}%
\bibitem [{\citenamefont {Roth}(2000)}]{libphd}%
  \BibitemOpen
  \bibfield  {author} {\bibinfo {author} {\bibfnamefont {E.~P.}\ \bibnamefont
  {Roth}},\ }\emph {\bibinfo {title} {Self-consistent Green's Functions in
  Nuclear Matter}},\ \href@noop {} {Ph.D. thesis},\ \bibinfo  {school}
  {Washington University in St. Louis} (\bibinfo {year} {2000})\BibitemShut
  {NoStop}%
\bibitem [{\citenamefont {Frick}(2004)}]{frickphd}%
  \BibitemOpen
  \bibfield  {author} {\bibinfo {author} {\bibfnamefont {T.}~\bibnamefont
  {Frick}},\ }\emph {\bibinfo {title} {Self-consistent Green's Functions in
  Nuclear Matter at Finite Temperature}},\ \href@noop {} {Ph.D. thesis},\
  \bibinfo  {school} {University of T{\"u}bingen} (\bibinfo {year}
  {2004})\BibitemShut {NoStop}%
\bibitem [{\citenamefont {Rios}(2007)}]{riosphd}%
  \BibitemOpen
  \bibfield  {author} {\bibinfo {author} {\bibfnamefont {A.}~\bibnamefont
  {Rios}},\ }\emph {\bibinfo {title} {Thermodynamical Properties of Nuclear
  Matter from a Self-Consistent Green's Function Approach}},\ \href@noop {}
  {Ph.D. thesis},\ \bibinfo  {school} {University of Barcelona} (\bibinfo
  {year} {2007})\BibitemShut {NoStop}%
\bibitem [{\citenamefont {Dickhoff}\ and\ \citenamefont {{Van
  Neck}}(2008)}]{Dickhoff2008}%
  \BibitemOpen
  \bibfield  {author} {\bibinfo {author} {\bibfnamefont {W.~H.}\ \bibnamefont
  {Dickhoff}}\ and\ \bibinfo {author} {\bibfnamefont {D.}~\bibnamefont {{Van
  Neck}}},\ }\href@noop {} {\emph {\bibinfo {title} {{Many-body theory
  exposed!}}}},\ \bibinfo {edition} {2nd}\ ed.\ (\bibinfo  {publisher} {World
  Scientific},\ \bibinfo {year} {2008})\BibitemShut {NoStop}%
\bibitem [{\citenamefont {Migdal}(1967)}]{Migdal1967}%
  \BibitemOpen
  \bibfield  {author} {\bibinfo {author} {\bibfnamefont {A.~B.}\ \bibnamefont
  {Migdal}},\ }\href@noop {} {\emph {\bibinfo {title} {{Theory of Finite Fermi
  Systems}}}}\ (\bibinfo  {publisher} {Interscience, New York},\ \bibinfo
  {year} {1967})\BibitemShut {NoStop}%
\bibitem [{\citenamefont {M\"{u}ther}\ and\ \citenamefont
  {Dickhoff}(2005)}]{Muther2005}%
  \BibitemOpen
  \bibfield  {author} {\bibinfo {author} {\bibfnamefont {H.}~\bibnamefont
  {M\"{u}ther}}\ and\ \bibinfo {author} {\bibfnamefont {W.~H.}\ \bibnamefont
  {Dickhoff}},\ }\href@noop {} {\bibfield  {journal} {\bibinfo  {journal}
  {Phys. Rev. C}\ }\textbf {\bibinfo {volume} {72}},\ \bibinfo {pages} {054313}
  (\bibinfo {year} {2005})}\BibitemShut {NoStop}%
\bibitem [{\citenamefont {Dickhoff}(1988)}]{Dickhoff88}%
  \BibitemOpen
  \bibfield  {author} {\bibinfo {author} {\bibfnamefont {W.~H.}\ \bibnamefont
  {Dickhoff}},\ }\href@noop {} {\bibfield  {journal} {\bibinfo  {journal}
  {Phys. Lett. B}\ }\textbf {\bibinfo {volume} {210}},\ \bibinfo {pages} {15}
  (\bibinfo {year} {1988})}\BibitemShut {NoStop}%
\bibitem [{\citenamefont {Dickhoff}\ and\ \citenamefont
  {M{\"u}ther}(1987)}]{Dickhoff1987}%
  \BibitemOpen
  \bibfield  {author} {\bibinfo {author} {\bibfnamefont {W.~H.}\ \bibnamefont
  {Dickhoff}}\ and\ \bibinfo {author} {\bibfnamefont {H.}~\bibnamefont
  {M{\"u}ther}},\ }\href@noop {} {\bibfield  {journal} {\bibinfo  {journal}
  {Nucl. Phys. A}\ }\textbf {\bibinfo {volume} {473}},\ \bibinfo {pages} {394}
  (\bibinfo {year} {1987})}\BibitemShut {NoStop}%
\bibitem [{\citenamefont {Cao}\ \emph {et~al.}(2006)\citenamefont {Cao},
  \citenamefont {Lombardo},\ and\ \citenamefont {Schuck}}]{Cao2006}%
  \BibitemOpen
  \bibfield  {author} {\bibinfo {author} {\bibfnamefont {L.~G.}\ \bibnamefont
  {Cao}}, \bibinfo {author} {\bibfnamefont {U.}~\bibnamefont {Lombardo}}, \
  and\ \bibinfo {author} {\bibfnamefont {P.}~\bibnamefont {Schuck}},\
  }\href@noop {} {\bibfield  {journal} {\bibinfo  {journal} {Phys. Rev. C}\
  }\textbf {\bibinfo {volume} {74}},\ \bibinfo {pages} {064301} (\bibinfo
  {year} {2006})}\BibitemShut {NoStop}%
\bibitem [{\citenamefont {Rios}\ \emph {et~al.}(2009)\citenamefont {Rios},
  \citenamefont {Polls},\ and\ \citenamefont {Dickhoff}}]{Rios2009a}%
  \BibitemOpen
  \bibfield  {author} {\bibinfo {author} {\bibfnamefont {A.}~\bibnamefont
  {Rios}}, \bibinfo {author} {\bibfnamefont {A.}~\bibnamefont {Polls}}, \ and\
  \bibinfo {author} {\bibfnamefont {W.~H.}\ \bibnamefont {Dickhoff}},\ }\href
  {\doibase 10.1103/PhysRevC.79.064308} {\bibfield  {journal} {\bibinfo
  {journal} {Phys. Rev. C}\ }\textbf {\bibinfo {volume} {79}},\ \bibinfo
  {pages} {064308} (\bibinfo {year} {2009})}\BibitemShut {NoStop}%
\bibitem [{\citenamefont {Rios}\ \emph {et~al.}(2014)\citenamefont {Rios},
  \citenamefont {Polls},\ and\ \citenamefont {Dickhoff}}]{Rios2014}%
  \BibitemOpen
  \bibfield  {author} {\bibinfo {author} {\bibfnamefont {A.}~\bibnamefont
  {Rios}}, \bibinfo {author} {\bibfnamefont {A.}~\bibnamefont {Polls}}, \ and\
  \bibinfo {author} {\bibfnamefont {W.~H.}\ \bibnamefont {Dickhoff}},\
  }\href@noop {} {\bibfield  {journal} {\bibinfo  {journal} {Phys. Rev. C}\
  }\textbf {\bibinfo {volume} {89}},\ \bibinfo {pages} {044303} (\bibinfo
  {year} {2014})}\BibitemShut {NoStop}%
\bibitem [{\citenamefont {Gorkov}(1958)}]{Gorkov58}%
  \BibitemOpen
  \bibfield  {author} {\bibinfo {author} {\bibfnamefont {L.~P.}\ \bibnamefont
  {Gorkov}},\ }\href@noop {} {\bibfield  {journal} {\bibinfo  {journal} {Sov.
  Phys.-JETP}\ }\textbf {\bibinfo {volume} {7}},\ \bibinfo {pages} {505}
  (\bibinfo {year} {1958})}\BibitemShut {NoStop}%
\bibitem [{\citenamefont {Som\`{a}}\ \emph {et~al.}(2011)\citenamefont
  {Som\`{a}}, \citenamefont {Duguet},\ and\ \citenamefont
  {Barbieri}}]{Barbieri2011}%
  \BibitemOpen
  \bibfield  {author} {\bibinfo {author} {\bibfnamefont {V.}~\bibnamefont
  {Som\`{a}}}, \bibinfo {author} {\bibfnamefont {T.}~\bibnamefont {Duguet}}, \
  and\ \bibinfo {author} {\bibfnamefont {C.}~\bibnamefont {Barbieri}},\
  }\href@noop {} {\bibfield  {journal} {\bibinfo  {journal} {Phys. Rev. C}\
  }\textbf {\bibinfo {volume} {84}},\ \bibinfo {pages} {064317} (\bibinfo
  {year} {2011})}\BibitemShut {NoStop}%
\bibitem [{\citenamefont {Bo{\.{z}}ek}(1999)}]{Bozek99}%
  \BibitemOpen
  \bibfield  {author} {\bibinfo {author} {\bibfnamefont {P.}~\bibnamefont
  {Bo{\.{z}}ek}},\ }\href@noop {} {\bibfield  {journal} {\bibinfo  {journal}
  {Nucl. Phys. A}\ }\textbf {\bibinfo {volume} {657}},\ \bibinfo {pages} {187}
  (\bibinfo {year} {1999})}\BibitemShut {NoStop}%
\bibitem [{\citenamefont {Dewulf}\ \emph {et~al.}(2003)\citenamefont {Dewulf},
  \citenamefont {Dickhoff}, \citenamefont {{Van Neck}}, \citenamefont
  {Stoddard},\ and\ \citenamefont {Waroquier}}]{Dewulf2003}%
  \BibitemOpen
  \bibfield  {author} {\bibinfo {author} {\bibfnamefont {Y.}~\bibnamefont
  {Dewulf}}, \bibinfo {author} {\bibfnamefont {W.~H.}\ \bibnamefont
  {Dickhoff}}, \bibinfo {author} {\bibfnamefont {D.}~\bibnamefont {{Van
  Neck}}}, \bibinfo {author} {\bibfnamefont {E.~R.}\ \bibnamefont {Stoddard}},
  \ and\ \bibinfo {author} {\bibfnamefont {M.}~\bibnamefont {Waroquier}},\
  }\href {\doibase 10.1103/PhysRevLett.90.152501} {\bibfield  {journal}
  {\bibinfo  {journal} {Phys. Rev. Lett.}\ }\textbf {\bibinfo {volume} {90}},\
  \bibinfo {pages} {152501} (\bibinfo {year} {2003})}\BibitemShut {NoStop}%
\bibitem [{Sup()}]{Suppl15}%
  \BibitemOpen
  \href@noop {} {}\bibinfo {note} {See Supplemental Material at
  http://link.aps.org/supplemental/10.1103/PhysRevLett.}\BibitemShut {Stop}%
\bibitem [{\citenamefont {Dickhoff}\ \emph {et~al.}(1982)\citenamefont
  {Dickhoff}, \citenamefont {Faessler},\ and\ \citenamefont
  {M{\"u}ther}}]{Dickhoff1982}%
  \BibitemOpen
  \bibfield  {author} {\bibinfo {author} {\bibfnamefont {W.~H.}\ \bibnamefont
  {Dickhoff}}, \bibinfo {author} {\bibfnamefont {A.}~\bibnamefont {Faessler}},
  \ and\ \bibinfo {author} {\bibfnamefont {H.}~\bibnamefont {M{\"u}ther}},\
  }\href@noop {} {\bibfield  {journal} {\bibinfo  {journal} {Phys. Rev. Lett.}\
  }\textbf {\bibinfo {volume} {49}},\ \bibinfo {pages} {1902} (\bibinfo {year}
  {1982})}\BibitemShut {NoStop}%
\bibitem [{\citenamefont {Pankratov}\ \emph {et~al.}(2015)\citenamefont
  {Pankratov}, \citenamefont {Baldo},\ and\ \citenamefont
  {Saperstein}}]{Pankratov2015}%
  \BibitemOpen
  \bibfield  {author} {\bibinfo {author} {\bibfnamefont {S.~S.}\ \bibnamefont
  {Pankratov}}, \bibinfo {author} {\bibfnamefont {M.}~\bibnamefont {Baldo}}, \
  and\ \bibinfo {author} {\bibfnamefont {E.~E.}\ \bibnamefont {Saperstein}},\
  }\href@noop {} {\bibfield  {journal} {\bibinfo  {journal} {Phys. Rev. C}\
  }\textbf {\bibinfo {volume} {91}},\ \bibinfo {pages} {015802} (\bibinfo
  {year} {2015})}\BibitemShut {NoStop}%
\bibitem [{\citenamefont {Friman}\ and\ \citenamefont
  {Dhar}(1979)}]{Friman1979}%
  \BibitemOpen
  \bibfield  {author} {\bibinfo {author} {\bibfnamefont {B.~L.}\ \bibnamefont
  {Friman}}\ and\ \bibinfo {author} {\bibfnamefont {A.~K.}\ \bibnamefont
  {Dhar}},\ }\href@noop {} {\bibfield  {journal} {\bibinfo  {journal} {Phys.
  Lett. B}\ }\textbf {\bibinfo {volume} {85}},\ \bibinfo {pages} {1} (\bibinfo
  {year} {1979})}\BibitemShut {NoStop}%
\bibitem [{\citenamefont {Machleidt}\ \emph {et~al.}(1996)\citenamefont
  {Machleidt}, \citenamefont {Sammarruca},\ and\ \citenamefont
  {Song}}]{Machleidt1995}%
  \BibitemOpen
  \bibfield  {author} {\bibinfo {author} {\bibfnamefont {R.}~\bibnamefont
  {Machleidt}}, \bibinfo {author} {\bibfnamefont {F.}~\bibnamefont
  {Sammarruca}}, \ and\ \bibinfo {author} {\bibfnamefont {Y.}~\bibnamefont
  {Song}},\ }\href {\doibase 10.1103/PhysRevC.53.R1483} {\bibfield  {journal}
  {\bibinfo  {journal} {Phys. Rev. C}\ }\textbf {\bibinfo {volume} {53}},\
  \bibinfo {pages} {R1483} (\bibinfo {year} {1996})}\BibitemShut {NoStop}%
\bibitem [{\citenamefont {Wiringa}\ \emph {et~al.}(1995)\citenamefont
  {Wiringa}, \citenamefont {Stoks},\ and\ \citenamefont
  {Schiavilla}}]{Wiringa1995}%
  \BibitemOpen
  \bibfield  {author} {\bibinfo {author} {\bibfnamefont {R.~B.}\ \bibnamefont
  {Wiringa}}, \bibinfo {author} {\bibfnamefont {V.~G.~J.}\ \bibnamefont
  {Stoks}}, \ and\ \bibinfo {author} {\bibfnamefont {R.}~\bibnamefont
  {Schiavilla}},\ }\href {\doibase 10.1103/PhysRevC.51.38} {\bibfield
  {journal} {\bibinfo  {journal} {Phys. Rev. C}\ }\textbf {\bibinfo {volume}
  {51}},\ \bibinfo {pages} {38} (\bibinfo {year} {1995})}\BibitemShut {NoStop}%
\bibitem [{\citenamefont {Bo{\.{z}}ek}(2000)}]{Bozek00}%
  \BibitemOpen
  \bibfield  {author} {\bibinfo {author} {\bibfnamefont {P.}~\bibnamefont
  {Bo{\.{z}}ek}},\ }\href@noop {} {\bibfield  {journal} {\bibinfo  {journal}
  {Phys. Rev. C}\ }\textbf {\bibinfo {volume} {62}},\ \bibinfo {pages} {054316}
  (\bibinfo {year} {2000})}\BibitemShut {NoStop}%
\bibitem [{\citenamefont {Baldo}\ and\ \citenamefont {Grasso}(2000)}]{Baldo00}%
  \BibitemOpen
  \bibfield  {author} {\bibinfo {author} {\bibfnamefont {M.}~\bibnamefont
  {Baldo}}\ and\ \bibinfo {author} {\bibfnamefont {A.}~\bibnamefont {Grasso}},\
  }\href@noop {} {\bibfield  {journal} {\bibinfo  {journal} {Phys. Lett. B}\
  }\textbf {\bibinfo {volume} {485}},\ \bibinfo {pages} {115} (\bibinfo {year}
  {2000})}\BibitemShut {NoStop}%
\bibitem [{\citenamefont {Baldo}\ \emph {et~al.}(2002)\citenamefont {Baldo},
  \citenamefont {Lombardo}, \citenamefont {Schulze},\ and\ \citenamefont
  {Wei}}]{Baldo02}%
  \BibitemOpen
  \bibfield  {author} {\bibinfo {author} {\bibfnamefont {M.}~\bibnamefont
  {Baldo}}, \bibinfo {author} {\bibfnamefont {U.}~\bibnamefont {Lombardo}},
  \bibinfo {author} {\bibfnamefont {H.~J.}\ \bibnamefont {Schulze}}, \ and\
  \bibinfo {author} {\bibfnamefont {Z.}~\bibnamefont {Wei}},\ }\href@noop {}
  {\bibfield  {journal} {\bibinfo  {journal} {Phys. Rev. C}\ }\textbf {\bibinfo
  {volume} {66}},\ \bibinfo {pages} {054304} (\bibinfo {year}
  {2002})}\BibitemShut {NoStop}%
\bibitem [{\citenamefont {Bo{\.{z}}ek}(2003)}]{Bozek03}%
  \BibitemOpen
  \bibfield  {author} {\bibinfo {author} {\bibfnamefont {P.}~\bibnamefont
  {Bo{\.{z}}ek}},\ }\href@noop {} {\bibfield  {journal} {\bibinfo  {journal}
  {Phys. Lett. B}\ }\textbf {\bibinfo {volume} {551}},\ \bibinfo {pages} {93}
  (\bibinfo {year} {2003})}\BibitemShut {NoStop}%
\bibitem [{\citenamefont {Entem}\ and\ \citenamefont
  {Machleidt}(2003)}]{Entem2003}%
  \BibitemOpen
  \bibfield  {author} {\bibinfo {author} {\bibfnamefont {D.~R.}\ \bibnamefont
  {Entem}}\ and\ \bibinfo {author} {\bibfnamefont {R.}~\bibnamefont
  {Machleidt}},\ }\href {\doibase 10.1103/PhysRevC.68.041001} {\bibfield
  {journal} {\bibinfo  {journal} {Phys. Rev. C}\ }\textbf {\bibinfo {volume}
  {68}},\ \bibinfo {pages} {041001} (\bibinfo {year} {2003})}\BibitemShut
  {NoStop}%
\bibitem [{\citenamefont {Ding}\ \emph {et~al.}(2014)\citenamefont {Ding},
  \citenamefont {Witte}, \citenamefont {Dickhoff}, \citenamefont {Dussan},
  \citenamefont {Rios},\ and\ \citenamefont {Polls}}]{Ding2014}%
  \BibitemOpen
  \bibfield  {author} {\bibinfo {author} {\bibfnamefont {D.}~\bibnamefont
  {Ding}}, \bibinfo {author} {\bibfnamefont {S.~J.}\ \bibnamefont {Witte}},
  \bibinfo {author} {\bibfnamefont {W.~H.}\ \bibnamefont {Dickhoff}}, \bibinfo
  {author} {\bibfnamefont {H.}~\bibnamefont {Dussan}}, \bibinfo {author}
  {\bibfnamefont {A.}~\bibnamefont {Rios}}, \ and\ \bibinfo {author}
  {\bibfnamefont {A.}~\bibnamefont {Polls}},\ }\href@noop {} {\bibfield
  {journal} {\bibinfo  {journal} {AIP Conf. Proc.}\ }\textbf {\bibinfo {volume}
  {1619}},\ \bibinfo {pages} {73} (\bibinfo {year} {2014})}\BibitemShut
  {NoStop}%
\bibitem [{\citenamefont {Maurizio}\ \emph {et~al.}(2014)\citenamefont
  {Maurizio}, \citenamefont {Holt},\ and\ \citenamefont
  {Finelli}}]{Maurizio2014}%
  \BibitemOpen
  \bibfield  {author} {\bibinfo {author} {\bibfnamefont {S.}~\bibnamefont
  {Maurizio}}, \bibinfo {author} {\bibfnamefont {J.~W.}\ \bibnamefont {Holt}},
  \ and\ \bibinfo {author} {\bibfnamefont {P.}~\bibnamefont {Finelli}},\
  }\href@noop {} {\bibfield  {journal} {\bibinfo  {journal} {Phys. Rev. C}\
  }\textbf {\bibinfo {volume} {90}},\ \bibinfo {pages} {044003} (\bibinfo
  {year} {2014})}\BibitemShut {NoStop}%
\bibitem [{\citenamefont {Sedrakian}(2003)}]{Sedrakian03}%
  \BibitemOpen
  \bibfield  {author} {\bibinfo {author} {\bibfnamefont {A.}~\bibnamefont
  {Sedrakian}},\ }\href@noop {} {\bibfield  {journal} {\bibinfo  {journal}
  {Phys. Rev. C}\ }\textbf {\bibinfo {volume} {68}},\ \bibinfo {pages} {065805}
  (\bibinfo {year} {2003})}\BibitemShut {NoStop}%
\bibitem [{\citenamefont {Carbone}\ \emph {et~al.}(2013)\citenamefont
  {Carbone}, \citenamefont {Polls},\ and\ \citenamefont {Rios}}]{Carbone2013a}%
  \BibitemOpen
  \bibfield  {author} {\bibinfo {author} {\bibfnamefont {A.}~\bibnamefont
  {Carbone}}, \bibinfo {author} {\bibfnamefont {A.}~\bibnamefont {Polls}}, \
  and\ \bibinfo {author} {\bibfnamefont {A.}~\bibnamefont {Rios}},\ }\href@noop
  {} {\bibfield  {journal} {\bibinfo  {journal} {Phys. Rev. C}\ }\textbf
  {\bibinfo {volume} {88}},\ \bibinfo {pages} {044302} (\bibinfo {year}
  {2013})}\BibitemShut {NoStop}%
\bibitem [{\citenamefont {Carbone}(2014)}]{carbonephd}%
  \BibitemOpen
  \bibfield  {author} {\bibinfo {author} {\bibfnamefont {A.}~\bibnamefont
  {Carbone}},\ }\emph {\bibinfo {title} {Self-consistent Green's Functions with
  Three-body Forces}},\ \href@noop {} {Ph.D. thesis},\ \bibinfo  {school}
  {University of Barcelona} (\bibinfo {year} {2014})\BibitemShut {NoStop}%
\bibitem [{\citenamefont {Dong}\ \emph {et~al.}(2013)\citenamefont {Dong},
  \citenamefont {Lombardo},\ and\ \citenamefont {Zuo}}]{Dong2013}%
  \BibitemOpen
  \bibfield  {author} {\bibinfo {author} {\bibfnamefont {J.~M.}\ \bibnamefont
  {Dong}}, \bibinfo {author} {\bibfnamefont {U.}~\bibnamefont {Lombardo}}, \
  and\ \bibinfo {author} {\bibfnamefont {W.}~\bibnamefont {Zuo}},\ }\href@noop
  {} {\bibfield  {journal} {\bibinfo  {journal} {Phys. Rev. C}\ }\textbf
  {\bibinfo {volume} {87}},\ \bibinfo {pages} {062801(R)} (\bibinfo {year}
  {2013})}\BibitemShut {NoStop}%
\end{thebibliography}%

\end{document}